\documentclass[journal,draftcls,onecolumn,12pt,twoside]{IEEEtran}

\usepackage{times,amsmath,epsfig,multirow}
\usepackage{algorithmic,algorithm}
\usepackage{epstopdf}
\usepackage{array}
\usepackage{color}
\usepackage{amssymb, mathrsfs}
\usepackage{graphicx}
\usepackage{url}
\usepackage{amsmath}
\usepackage{cite}
\usepackage{textcomp}
\usepackage{soul}
\usepackage{microtype,tabularx,booktabs}
\usepackage[labelformat=simple]{subfig}

\def\ve#1{{\mathchoice{\mbox{\boldmath$\displaystyle #1$}}%
{\mbox{\boldmath$\textstyle #1$}}%
{\mbox{\boldmath$\scriptstyle #1$}}%
{\mbox{\boldmath$\scriptscriptstyle #1$}}}}

\begin{document}

\title{Power Line Communication and Sensing Using Time Series Forecasting}

\author{Yinjia~Huo,~\IEEEmembership{Student~Member,~IEEE,}
	Gautham~Prasad,
       Lutz~Lampe,~\IEEEmembership{Senior Member,~IEEE,}  
       and Victor~C.~M.~Leung,~\IEEEmembership{Life Fellow,~IEEE}
       \thanks{Yinjia Huo is with the Department of Electrical and Computer Engineering, The University of British Columbia, Vancouver, BC, Canada. Gautham Prasad is with Ofinno, Reston, VA, USA. Lutz Lampe is with the Department of Electrical and Computer Engineering, The University of British Columbia, Vancouver, BC, Canada. Victor C.M. Leung is with the College of Computer Science and Software Engineering, Shenzhen University, Shenzhen, Guangdong, China. Email: yortka@ece.ubc.ca, gautham.prasad@alumni.ubc.ca, lampe@ece.ubc.ca, vleung@ieee.org.}
      }%

\maketitle

\begin{abstract}
Smart electrical grids rely on data communication to support their operation and on sensing for diagnostics and maintenance. Usually, the roles of communication and sensing equipment are different, i.e., communication equipment does not participate in sensing tasks and vice versa. Power line communication (PLC) offers a~cost-effective solution for \textsl{{joint}} communication and sensing for smart grids. This~is because the~high-frequency PLC signals used for data communication also reveal detailed information regarding the~health of the~power lines that they travel through. 
  Traditional PLC-based power line or cable diagnostic solutions are dependent on prior knowledge of the~cable type, network topology, and/or characteristics of the~anomalies. In~this paper, we~develop a~power line sensing technique that can detect various types of cable anomalies without any prior domain knowledge. To~this end, we~design a~solution that first uses time-series forecasting to predict the~PLC channel state information at any given point in~time based on its historical data. Under the~approximation that the~prediction error follows a~Gaussian distribution, we~then perform chi-squared statistical test to build an~anomaly detector which identifies the~occurrence of a~cable fault. We~demonstrate the~effectiveness and universality of our sensing solution via evaluations conducted using both synthetic and real-world data extracted from low- and medium-voltage distribution networks.
  \end{abstract}


\IEEEpeerreviewmaketitle

\section{Introduction}
\label{sec:introduction}

\subsection{Background}
\label{subsec:motivate}

Asset monitoring is critical for the~safe and smooth operation of the~electricity grid system~\cite{farhangi2010path}. The~advent of the smart grid, which allows for bidirectional data exchange between the~utility and the~consumer~\cite{sg2013, gungor2011smart, fang2011smart}, unfolds a~new paradigm of solutions for smart grid sensing and infrastructure monitoring to improve~system resilience of the~grid. Among~various smart grid data communication solutions, power line communications (PLC) use the~existing power line infrastructure, which was originally designed to transport electric energy and for~communication purposes. In~this paper, we~further propose to re-use PLC modems for cable diagnostics as~a~use case of joint communication and sensing~\cite{wild2021joint} for the~smart grid. This~provides the~benefits of realizing a~low-cost solution that can operate in~an online, independent, and~automatic manner without requiring any new component installations~\cite{huo2019cable, prasad2019fault, lehmann2016diagnostic, passerini2017full, passerini2017analysis, prasad2019full}.

PLC is a~commonly used solution to enable information and communication technology for the~smart grid~\cite{galli2011grid, openriva, mengi2017itut}. {Power line modems (PLMs) that transmit and receive smart grid data constantly estimate the~power line channel state information (PLCSI) for adapting their operation. In~this context, we~refer to PLCSI as any parameter that conveys the~channel behavior, either directly, e.g.,~channel frequency response (CFR) or access impedance, or~indirectly, e.g.,~signal-to-noise ratio (SNR) or precoder matrix.} Prior articles  
have shown that this estimated PLCSI also contains information that can be used as~a~smart grid sensing solution to infer cable health conditions~\cite{lehmann2016diagnostic, passerini2017full, passerini2017analysis, huo2019cable, prasad2019fault, prasad2019full}. We~build on these methods to propose an~enhanced cable diagnostics solution using time series~forecasting.

\subsection{Related~Works}
\label{subsec:literature}
To learn from and contrast our proposed solution to prior articles, 
we~present a~brief overview of the~existing body of literature in~the~broad area of diagnostics by categorizing them into the~following~topics. 
\subsubsection{Legacy Cable~Diagnostics}
Some legacy cable diagnostic solutions, e.g.,~reflectometry-based methods, require deployment of specialized equipment and/or personnel to conduct the~tests~\cite{wang2010application, dubickas2006line}, \mbox{{(}\cite{gill2008electrical} {Ch. 6)},}~\cite{furse2020fault}. Furthermore, several non-PLC solutions that sample the~electrical signal with a~lower frequency, such as phasor measurement units, suffer from noisy data impacted by electrical disturbances, and~are unable to discern precise information about cable defects, as~seen in, e.g.,~\cite{farajollahi2018locating} to determine the~age of degradation or accurate location of the~position of a~fault~\cite{dubickas2006line, farajollahi2018locating}. Our proposed PLC-based monitoring technique, on~the other hand, reuses the~high-frequency broadband communication signals as probing waves to provide effective cable diagnostics similar to other PLC-based diagnostic solutions as presented in~\cite{huo2018cable, huo2019cable}.

\subsubsection{PLC Cable Diagnostics~Solution}
Various PLC-based solutions have been developed in~the~past to exploit the~\textit{{through-the-grid}} nature of PLC signals to gain insights into the~characteristics of the~transmission medium~\cite{lehmann2016diagnostic, passerini2017full, passerini2017analysis, huo2019cable, prasad2019fault, prasad2019full, alam2012design}. Many of the~proposed PLC-based diagnostic solutions typically require a~reference healthy measurement, i.e.,~PLCSI of the~exact type of the~deployed cable that is undamaged (e.g.,~\cite{lehmann2016diagnostic, passerini2017full, passerini2017analysis}). PLCSI estimated within~the~PLM is then compared against this reference measurement to infer the~health of the~cable. Aside from practical challenges associated with employing such a~method in~the~real world, these methods also provide less than reliable results in~anomaly detection since the~load conditions are constantly fluctuating, 
which renders it~hard to distinguish between benign and malicious PLCSI changes, e.g.,~those that are caused due to load variations as opposed to grid anomalies. Alternatively, data-driven methods that were designed to use machine-learning (ML) techniques to intelligently detect and assess cable health are resilient against such challenges~\cite{huo2019cable, prasad2019fault, huo2019smart}. These~methods harness ML classification and regression techniques to detect, locate, and~assess various smart grid network anomalies, such as cable degradation and faults and network intrusions~\cite{huo2019cable, huo2018cable, prasad2019fault, prasad2019machine}. One notable work is~\cite{bondorf2021broadband}, which extracts features from the~PLM SNR spectrogram and conducts unsupervised clustering for anomaly detection. However, these methods are not universally applicable since the~machines used here are typically trained under a~specific operating network topology to detect a~few known types of characterized anomalies. When the~machine is deployed under a~different network topology or is applied to detect a~type of anomaly it~has never encountered in~the~process of training, the~performance of these solutions suffer significantly. Our proposed universal cable anomaly detector counters these drawbacks by processing the~collected PLC signals and/or PLCSI data using time-series forecasting to determine a~variety of anomalies in~a~smart grid network. Unlike the~above methods, our solution does not demand stringent machine training regimens that are heavily reliant on~the~accuracy of the~network topology emulation or load~modeling.

\subsubsection{Time-Series Prediction for~Diagnostics}
Our proposed solution includes a~two-pronged approach with a~time-series forecasting followed by the~anomaly detection procedure. Prior articles 
on~the~former use auto-regressive integrated moving-average (ARIMA) and NN-based forecasting to provide accurate prediction results for a~variety of applications such as gearbox fault trend prediction~\cite{yang2018application}, peak load value forecasting~\cite{wang2018high}, and~other seasonal and time-series trend analyses~\cite{bandara2020forecasting, khandelwal2015time, zhang2005neural, allende2002artificial}. The~issue with these prior articles 
is that they fall short of the~task that we~consider, i.e.,~detecting anomalies using the~predicted data. We~nevertheless borrow approaches from the~aforementioned works for our first task of time-series forecasting, and~then further build on~the~work to use the~results from the~prediction to determine possible anomalies by comparing them against practical values obtained from a~device under test. To~this end, forecasting approaches developed in~the~past form a~solid starting point for our proposed~work.

\subsubsection{Cable Diagnostics Solution Using Other~Techniques}
\textls[-15]{Several cable diagnostic solutions have been historically proposed that use high-frequency probing or testing signals, for~example, using time-frequency reflectometry~\cite{lee2018time, bang2019classification, kwon2017offline, lee2018monitoring, lee2018multi, de2019orthogonal}. Such techniques have been shown to be efficient at detecting abrupt cable anomalies, e.g.,~a cable fault introduced due to neighboring infrastructure upgrade, such as a~neighboring line drilling. However, these methods are not suitable for diagnosing cable anomalies that progressively develop over a~longer period of time. For~instance, some incipient faults do not typically manifest as distinct changes in~waveform responses observed between relatively closer instances of time, and~could instead take a~considerable amount of time to be seen as~a~noticeable change in~the~probing waveform. Our solution is applicable to both these types of faults and can also be used for capturing anomalies that tend to manifest over a~longer duration of~time.}

\subsubsection{Diagnostics Solution for Other~Systems}
A vast body of literature exists in~the~broad domain of diagnostics that neither use time-series forecasting-based techniques nor are targeted at detecting anomalies for cables. The~works in~\cite{nabeshima2005line, catterson2009line, lin2018method, ashwini2017wireless, wang2014fault} are some of the~classical works to name a~few. While we~recognize the existence of these prior articles 
that focus on a problem similar to that of ours, techniques of little relevance are found in~these works that can be used or adapted to counter the~challenges we~face in~our considered setup. We~highlight that most of these works are predominantly suitable for detecting, locating, and/or assessing \textit{{abrupt}} faults that manifest as noticeable response change in~a~short duration time. In~contrast, our time-series prediction-based diagnostics solution is a~powerful method to detect a~host of anomalies of varied nature without requiring prior domain knowledge of the~infrastructure under~test.

\subsection{Contributions}
\label{subsec:contribute}

In this paper, we~develop a~general purpose cable anomaly detector. This~means our developed cable anomaly detector does not require any reference measurements from healthy cables and is universally applicable. Our design is fully agnostic to the~nature of the~anomaly, i.e.,~its physical behavior and its cause, and~to the~infrastructure configuration, such as cable type or network topology. To~this end, we~propose the~use of historical PLCSI{, in~particular the~SNR as we~will explain below, for~the link between a~transmitting and a~receiving PLM}, to~train~a~time-series predictor. By~treating the~time-stamped PLCSI as time-series data, we~use time-series forecasting to predict the~PLCSI at any given point in~time using historical data by exploiting the~knowledge that the~network topology, cable configuration, and~the physical properties of the~cable are relatively stable for extended periods of time. In~addition, since the~long-term load {(throughout this paper, we~refer to \textit{load} as the~impedance load connected to a~circuitry, as~opposed to the~electricity consumption load of a~grid)} conditions are closely related to their historical values, the~PLCSI is also correlated in~time and can be predicted using historical state information~\cite{hagan1987time}. We~then compare the~predicted response against the~actual response estimated by the~PLM to detect a~potential~anomaly.

The performance of our solution relies heavily on~the~accuracy of the~predicted PLCSI. With~a highly accurate prediction, the~detector would be capable of detecting even subtle faults, which might not be discernible if~the~prediction itself is noisy. To~this end, we~investigate a~range of possible candidates for forecasting, including classical approaches such as the~ARIMA model~{(}\cite{box2011time}, {Ch.~4)} and feed-forward neural networks (FFNNs)~\cite{allende2002artificial}, and~also relatively recently developed techniques such as the long short-term memory (LSTM) model~\cite{malhotra2015long}. Furthermore, owing to its success in~previous PLC-based cable diagnostics~\cite{huo2018cable, huo2019cable}, we~also evaluate the~use of least-square boosting (L2Boost)~\cite{schapire2012boosting}.

The second factor of consideration toward building our solution is the~design of the~cable anomaly detector based on~the~predicted and the~measured PLCSI values. PLMs estimate a~range of { instances of PLCSI for adapting data communication} in~a~time varying environment. Some of the~estimated PLCSIs that shed direct light on~the~channel and in~turn on~the~cable health are the~end-to-end CFR, access impedance, precoder matrix, and~self-interference channel impulse response~\cite{yonge2013overview, prasad2016band, prasad2017digitally}. {However, several existing PLM chip-sets are unable to extract these parameters in~their entirety without additional firmware modifications~\cite{prasad2019fault}. In~light of this, we~consider the~use of SNRs instead, which can be readily extracted from current-day PLM chip-sets~\cite{prasad2019fault} and can be used for processing either locally within~the~PLM or reported to a~common location by all PLMs, e.g.,~a sub-station, for~centralized data processing.} The challenge lies in~differentiating between a~cable anomaly and an~inaccurate prediction. Furthermore, the~use of SNRs as PLCSIs introduces an~added challenge of designing a~cable diagnostics solution with poor quality information, since SNRs contain lesser insight into the~power line medium when compared to, e.g.,~CFR, together with being distorted by ambient noise. To~counter these impacts, we~exploit the~orthogonal frequency-division multiplexing (OFDM) nature of broadband PLC transmissions~\cite{lampe2016power}. We~first divide all the~OFDM subcarriers into several groups and average the~value of SNRs across all subcarriers within each group. This~stabilizes the~group SNR average, which then in~turn also makes it~more accurately predictable. With~the approximation that the~prediction errors across the~subcarrier groups follow a~multi-variate Gaussian distribution as in~\cite{o2016recurrent, cui2018anomaly}, we~determine a~probable occurrence of an~anomaly event based on~the~significance level of the~squared Mahalanobis distance (SMD){~\cite{slotani1964tolerance}}. The~significance level can be determined either empirically from the~training data or theoretically from a~chi-squared test~\cite{plackett1983karl}.

We verify the~feasibility and the~effectiveness of our proposed schemes through numerical evaluations using both synthetic data and in-field collected data. For~the former, we~use a~bottom-up PLC channel emulator to generate the~{SNR} time-series data, which allows us to investigate the~performance of our proposed solution under various types of cable anomalies in~a~customized and a~controlled environment. The~in-field data obtained from~\cite{hopfer2020nutzen} further allows us to verify our proposed schemes in~the~real-world, which indicates the~performance of our proposed technique in~practice.

To summarize, in~this work, we~propose a~general purpose cable anomaly detector using improvised grid sensors, i.e.,~PLMs. Our proposed solution does not require any prior domain knowledge of the~infrastructure under test and is universally applicable. We~design our anomaly detection schemes based on~the~techniques of time-series forecasting and the~statistical test of prediction errors. We~verify the~feasibility and effectiveness of our proposed schemes through numerical evaluations using both synthetic  and in-field collected~data.

\subsection{Paper~Organization}
\label{subsec:organize}

In Section~\ref{sec:timeSeriesPrediction}, we~formulate our time-series forecasting problem and elaborate our techniques for the~time-series prediction. Further, in~Section~\ref{sec:anomalyDetection}, we~introduce our cable anomaly detection scheme based on~the~time-series prediction results, including the~construction of stabilizer group and the~detection based on~the~use of SMD for the~chi-squared test. We~present our findings from our case studies of our proposed schemes in~Section~\ref{sec:caseStudy}. Supplementary analyses, including a~robustness test and discussion of incipient faults, are provided in~Section~\ref{sec:discuss}. Section~\ref{sec:conclusion} concludes the~paper.

\section{Time-Series~Forecasting} 
\label{sec:timeSeriesPrediction}
We begin by presenting a~brief overview of time-series prediction by focusing on~the~pertinent algorithms that we~consider for our proposed method. This~helps us in~understanding the~performance of the~{SNR} forecasting using time-series data. {We note that the~SNR for the~link between two PLMs that is measured in~the~receiving PLM is the~ratio of the~power of the~received data signal of interest to the~power of the~background noise. As~we will discuss in~Section~\ref{subsec:groupSNR}, the~SNR is typically measured for each OFDM subcarrier in~a~PLM. Since this is not significant for the~discussion of the~prediction algorithms,  we~omit the~subcarrier index for the~moment.}

\subsection{Time-Series Data for Cable Anomaly~Detection}
\label{subsec:predictionFormulation}

The time-stamped SNR between a~transmitter--receiver PLM pair is denoted as $x_j$, where $j$ is the~integer discrete time index. We~formulate our problem as using windowed instances of $x_j$, where $n-w \leq j<n$, to~predict $x_n$ and obtain~the~predicted value as $\tilde{x}_n$, with~$w$ being the~window size. {While a~larger $w$ can improve the~prediction performance, we~note that increasing $w$ requires more training data to determine the~predictor parameters, leads to a~higher computational complexity for prediction, and~only provides  incremental benefits beyond a~certain point. Hence, the~value of $w$ should thus be determined by considering the~trade-off between benefits and costs.}

Among the~available samples of $x_j$, we~use $x_j$, where $j \leq n_\text{tr}$, to~train~the~time-series predictor, where $n_\text{tr}$ is the~number of samples used for training. Once the~model is trained, we~then use it~to predict $\tilde{x}_j$, where $j > n_\text{tr}$. We~use the~normalized root mean square error (RMSE), $\eta$, as~the performance indicator of our prediction, which is computed as
\begin{equation}
    \eta = {\frac{\sqrt{\sum \limits_{j = n_\text{tr}+1}^{N} (x_j - \tilde{x}_j)^2 }} {\sqrt{\sum \limits_{j = n_\text{tr}+1}^{N} (x_j - \mu_x)^2} }},
\end{equation}
where $\mu_x$ is the~sample mean of the~observations of $x_j$ for $n_\text{tr}+1 \leq j \leq N$, and~$N$ is the~total number of $x_j$ samples used for training and testing. {Typically, we~would want a~large value of $N$, so as to saturate the~training of the~predictors' parameters and to reliably measure their performances.}

To compare the~performance of our ARIMA and ML-based predictors against a~baseline approach, we~consider a~simple extrapolation,
\begin{equation}\label{eq:baseline}
    \tilde{x}_{n} = x_{n-1}.
\end{equation}

In the~following, we~discuss the~use of different time-series forecasting methods for predicting $\tilde{x}_n$. We~defer to Section~\ref{subsec:predictDataSet} for the~procedure to choose suitable time-series prediction models to use for our anomaly detection, depending on~the~nature of the~data used for our diagnostics~scheme.

\subsection{ARIMA}
\label{subsec:predictionARIMA}

The ARIMA model is a~classical time-series predictor that has successfully been used across various domains of application, including financial series prediction, demand and load prediction in~the~power generation and distribution industry, and~customer sales prediction~{(}\cite{box2011time}, {Ch.~1)}. An~ARIMA model is specified by its order and its associated parameters. A~$(p,d,q)$ ARIMA model is a~$p$th order auto-regressive, $q$th order moving-average linear model with $d$th order of difference. A~$(p,d,q)$ ARIMA model has $p$ auto-regressive terms with $p$ auto-regressive coefficients and $q$ moving-average terms with $q$ moving-average coefficients. A~$d$th order difference is generated using $d$ subtraction operations, i.e.,~$u_{d,j}=u_{d-1,j}-u_{d-1,j-1}$ for $d\geq2$ with $u_{1,j}=x_{j}-x_{j-1}$.

The resultant time-series after difference is then assumed to be a~$(p,q)$ auto-regressive moving-average model, which is a~linear model with $p$ auto-regressive terms and $q$ moving-average terms, which is specified by
\begin{equation}
\label{eqn:DefARMA}
{u_{d,j} = \sum \limits_{i=1}^{p}\phi_i u_{d,j-i}+a_{j}-\sum \limits_{i=1}^{q}\theta_i a_{j-i},}
\end{equation}
where $\phi_i$ are coefficients for auto-regressive terms, $\theta_i$ are coefficients for moving-average terms, and~$a_j$ is the~random shock terms drawn independently from a~Gaussian distribution having zero mean and variance $\sigma_{a}^2$.

\subsection{Least-Square~Boosting}
\label{subsec:predictionL2Boost}
As our second time-series predictor candidate, we~investigate L2Boost, which has been shown to be successful in~the~past, specifically for cable diagnostics~\cite{huo2018cable, huo2019cable}. L2Boost is a~popular ML technique used for supervised regression tasks~\cite{schapire2012boosting}. It~is one of the~meta-ML algorithms which works by consolidating multiple weak learners into a~strong learner~\cite{freund1999short}. It~applies the~weak learners sequentially to weighted versions of the~data, where a~higher weight is allocated to examples that suffered greater inaccuracy in~earlier prediction rounds. These~weak learners are typically only marginally better than random guessing but are computationally simple. Boosting is also known to be robust to over-fitting, and~can be efficiently executed since it~is a~forward stage-wise additive~model.

To use the~L2Boost for time-series prediction, we~organize the~SNR time series into a~labeled data set for the~supervised learning. For~the training data set, i.e.,~$x_j$, where $1 \leq j \leq n_\text{tr}$, we~prepare each sample with {input} 
 $\ve{x}_j=(x_j,x_{j+1},\dots,x_{j+w-1})$ and its associated label $y_j=x_{j+w}$, where $j+w \leq n_\text{tr}$. We~then prepare the~testing samples in~a~similar way with input from $x_j$ to $x_{j+w-1}$ and its associated label as $x_{j+w}$, but~with {$j > n_\text{tr}$}.

\subsection{Feed-Forward Neural Network and~Long-Short-Term-Memory}
\label{subsec:predictionNN}
As our last set of predictor candidates, we~investigate the~use of two types of artificial neural network (ANN) models, FFNN and LSTM. Despite the~absence of feature engineering, ANNs can still explore the~inherent structure of the~input data, which could be hidden and/or complex. The~architecture of ANN is flexible with varying number of hidden layers and neurons in~each layer. To~use ANNs for time-series prediction, we~organize the~PLCSI values into a~labeled data {set of} $\ve{x}_j$ and $y_j$ for the~supervised learning the~same manner as in~Section~\ref{subsec:predictionL2Boost}.

While the~FFNN has~a~plain architecture, where the~output of the~previous layer is fed as the~input to the~current layer, i.e.,~feed-forward from the~input layer to the~output layer, the~LSTM has~a~feed-back mechanism, where the~output of the~current layer at the~last time stamp together with the~output of the~previous layer at the~current time stamp are fed as the~input to the~current layer at the~current time stamp. For~the LSTM model, the~feed-back of the~current layer from the~last time stamp is controlled by a~forgetting gate and the~output of the~previous layer at the~current time stamp is controlled by an~input gate. The~forgetting gate controls how much previous information memorized by the~LSTM machine is forgotten and the~input gate controls how much new information from the~input layer is passed through the~LSTM machine. Such a~feed-back mechanism is capable of capturing a long-term time dependence relationship and suitable for a~variety of time-series prediction tasks. When such a long-term time dependence relationship is not present, using FFNN in~place of an~LSTM machine can reduce the~risk of~over-fitting.

\section{Cable Anomaly~Detection} 
\label{sec:anomalyDetection}

{In this section, we~present the~design of the~cable anomaly detector. We~consider PLMs deployed in~a~smart grid that communicate with each other through the~grid's power lines. 
  For a~link between two PLMs, which could be part of a~longer multi-hop PLC connection, the~receiving PLM measures the~SNR. The~current SNR measurement and its predicted value, which are obtained from previous measurements and time-series forecasting as described in~Section~\ref{sec:timeSeriesPrediction}, are input to the~anomaly detector. An~anomaly located between the~two PLMs is identified by comparing the~difference between the~measured and the~predicted SNR against a~threshold, as~will be explained in~more detail below.  The~objective of the~anomaly detector is to maximize the~detection rate while simultaneously not exceeding a~set probability of false alarm (FA).

  We note that we~use the~term ``cable anomaly'' fairly broadly, as~there are various factors that affect the~communication quality as quantified by the~SNR. First, while our work, and~in particular our numerical results presented in~Section~\ref{sec:caseStudy}, focus on underground cables, the~presented concept is also applicable to overhead power lines. Second, since the~PLC signal is actively probing the~power line infrastructure, our method is able to detect anomalies resulting from physical degeneration of cables. We~will describe several models for such a~degeneration in~Section~\ref{subsec:detectDataSet}. However, anomalies in~other infrastructure elements such as cable joints or fuses would also affect the~SNR, as~well as changes in~the~noise environment due to, e.g.,\ imminent equipment failures.}



\subsection{Data~Preparation} 
\label{subsec:groupSNR}


{For the~overwhelming majority of broadband PLC transceivers that use OFDM transmission~\cite{lampe2016power}, SNRs are measured individually at each subcarrier. To~stabilize the~time-series SNR data,} we~divide all the~OFDM subcarriers into multiple groups called \textit{{stabilizer batches}}. We~then average the~SNR across all individual subcarriers within each stabilizer batch. This~procedure of averaging within~a~stabilizer batch ensures that the~time-series SNR data {are more predictable when compared to using SNRs of individual subcarriers.} In this regard, it~is essential to have the~subcarriers within~a~batch to be contiguous. This~ensures that, within~each stabilizer batch, the~variations in~individual SNR values are only gradual and the~impacts of cable anomalies on~the~individual subcarrier SNRs are similar in~nature.

This process results in~several stabilizer batches, and~the time-stamped average SNR values in~each individual stabilizer batch are treated as~a~set of time-series data. We~denote $\ve{z}_i = \{z_{i,j}\}$, $1 \leq i \leq n_\text{SB}$ to denote the~time series of the~average SNR of the~$i$th stabilizer batch, where $n_\text{SB}$ is the~number of such stabilizer batches. For~every $i$th stabilizer batch, we~use the~candidate forecasting models described in~Section~\ref{sec:timeSeriesPrediction} to develop a~time-series predictor $F_i$ to predict the~average time-series SNR, $\tilde{\gamma}_{i,j}$. The~input to the~predictor is the~windowed time series
\begin{equation}
\ve{v}_{i,j}=[z_{i,j},z_{i,j+1},\dots,z_{i,j+w-1}]^\mathrm{T},
\end{equation} 
with the~samples corresponding to $j+w \leq n_\text{tr}$ used during training and those corresponding to $j+w > n_\text{tr}$ used while testing. Hence, the~prediction is $\tilde{\gamma}_{i,j}=F_i(\ve{v}_{i,j})$ while the~true label is $\gamma_{i,j}=z_{i,j+w}$.

\subsection{Detection Using Squared Mahalanobis~Distance} 
\label{subsec:synthesizeSNR}
To detect an~anomaly we~consider the~difference in~the~predicted SNR to the~one measured by the~PLM,
\begin{equation}
    \delta_{i,j}=\tilde{\gamma}_{i,j}-\gamma_{i,j}. 
\end{equation}

{ The measured SNR is a~random variable affected by several aspects in~the~grid such as loads that are  random in~nature. The~prediction error is a~result of the~linear (ARIMA) or non-linear (L2Boost, FFNN, LSTM) superposition of several such measured SNRs.}  Therefore, we~approximate $\delta_{i,j}$ as~a~multi-variate Gaussian distribution as in~\cite{o2016recurrent, cui2018anomaly}, which is stationary over $j$, with~mean $\ve{\mu}$ and covariance matrix $\ve{\Sigma}$. With~$\ve{\delta}_{j}=[\delta_{1,j}, \delta_{2,j}, ..., \delta_{n_\text{SB},j}]^\mathrm{T}$, we~compute the~SMD as
\begin{equation}\label{def_dma}
D_\text{MA}^2=(\ve{\delta}_{j}-\ve{\mu})^{T} \ve{\Sigma}^{-1} (\ve{\delta}_{j}-\ve{\mu}).
\end{equation}

\textls[-25]{$D_\text{MA}^2$ follows a~chi-squared distribution with a~degree of freedom of $\kappa=n_\text{SB}$. Then, following the~theory of chi-squared statistical test~\cite{plackett1983karl}, for~a significance level of $\alpha$, we~define the~quantile function of the~chi-squared distribution with a~degree of freedom $\kappa$, as~$\chi_\kappa^2(\cdot)$, i.e.,}
\begin{equation}
    \mathsf{Pr}(D_\text{MA}^2 \leq \chi_\kappa^2(1-\alpha)) = 1 - \alpha,
\end{equation}
where $\mathsf{Pr}(\cdot)$ is the~probability function. Finally, for~a chosen {target FA rate} of $p_\text{FA}$, our anomaly detector declares a~warning of a~potential cable anomaly when
\begin{equation}
    D_\text{MA}^2 > T_{\text{r}}(p_\text{FA}),
\end{equation}
where the~threshold $ T_{\text{r}}(p_\text{FA})$ is determined according to the~corresponding significance level by
\begin{equation}\label{eq_threshold}
T_{\text{r}}(p_\text{FA})=\chi_\kappa^2(1-p_\text{FA}).
\end{equation}

\section{Design and Case~Studies} 
\label{sec:caseStudy}

We now highlight the~performance of our proposed cable anomaly detection by applying it~to two different types of data sets, one generated synthetically and the~other collected in-field, and~we describe the~design details~involved.

\subsection{Data~Sets} 
\label{subsec:dataSet}
\subsubsection*{In-Field~Data}
We acquire in-field measurements from the~data made available to us by the~author of~\cite{hopfer2020nutzen}. The~data were measured using 24 BB-PLC modems installed in~the~low-voltage (LV) sector of a~distribution network and 12 BB-PLC modems in~the~medium-voltage (MV)~grid. {The PLMs in~the~LV network were placed at house connection and at distribution points, and~several point-to-multi-point connections were possible. Overall these formed 22 transmitter--receiver pairs with 44 bidirectional data transmission links. The~PLMs in~the~MV network established point-to-point connections, and~we thus have 6 pairs with 12 PLC links.  The~SNR data for PLC-links were measured by the~receiving PLMs} every 15~min over $917$ OFDM subcarriers spaced $24.414$~kHz apart. This~data collection spanned an~overall time period from $17$ to $21$ months.

Due to limitations in~generating flexible observations and anomalies in~practical grids, the~in-field data consist of only two recorded instances of network anomalies. Furthermore, although~information of the~cable type, length, and~the biological age of the~cables are provided in~\cite{hopfer2020nutzen}, there is limited information available on~the~operation condition during the~field test. Therefore, for~a comprehensive evaluation, together with using the~{in-field collected data}, we~also use synthetic data sets obtained from constructing a~PLC network and generating {SNR time-series data} using the~bottom-up~approach. 

\begin{figure}[H]
	\includegraphics[width=10cm]{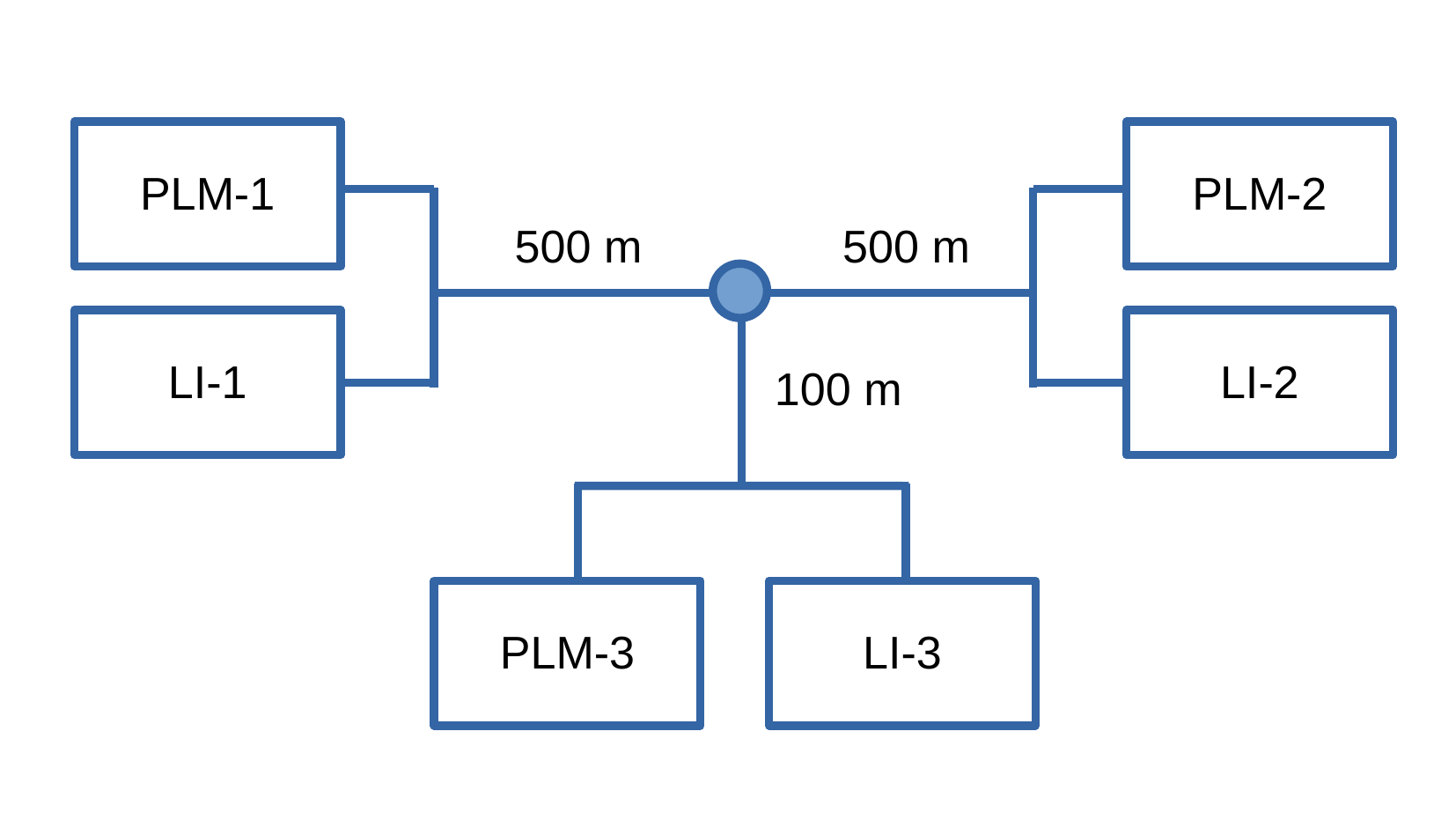}
	\caption{{PLC network} topology for synthetic data generation.}
	\label{fig:PLCTnetwork}
\end{figure}

For consistency between the~two types of data sets, we~borrow several network settings for generating the~synthetic data from the~in-field measurement campaign. We~generate { samples for the~SNR} between a~pair of PLMs for every $15$ min over a~period of $664$ days. { For this, we~adopt the~same OFDM parameters as listed above, a~transmit power spectral density (PSD) of $-$50~dBm/Hz, 
 and a~noise PSD of  $-$120~dBm/Hz, which are typical values for PLC. The~channel is generated with an~{emulator~\cite{gruber2015plc}} for the~constructed PLC network, which we~choose as~a~T-topology as shown in~Figure~\ref{fig:PLCTnetwork}. The~SNR measurements are performed at PLM-2 for the~link between PLM-1 and PLM-2, i.e.,~anomalies anywhere along the~hop from PLM-1 to PLM-2 including in~the~branch leading to PLM-3 are targeted.  The~three PLMs are connected through multi-core N2XSEY HELUKABEL cables with cross-linked polyethylene (XLPE) insulation, whose configuration and parameters can be found in~{(}\cite{forstel2017grid}{,~Table~2)}. Load impedances (LIs) which represent the~power grid extending outward from the~T-topology are connected in~parallel to the~PLMs. That is, the~LIs are the~equivalent impedances of the~surrounding grid experienced at the~edges of the~T-topology. } 

We consider three types of time-series LI models to emulate the~temporal dependence of electrical loads, motivated by seasonal and auto-regressive properties of loads in~the~mains frequency~\cite{hagan1987time}. We~denote the~load value at discrete time index $j$ of the~LI model $k$, $k\in\{1,\,2,\,3\}$, as~$L_{k,j}$. For~$k=1,\,2$, we~apply a~second-order auto-regressive model and a~cyclic model with one day per cycle, respectively. Furthermore, we~add \textit{{random shocks}}, $r_{k,j}$, to~the models to introduce a~degree of randomness in~the~load variations. As~a result, our first LI model is
\begin{align}
    L_{1,j} = 
    \begin{cases}
      r_{1,1}, & \text{for } j=1 \\
      0.8L_{1,1} + r_{1,2}, & \text{for } j=2 \\
      0.6L_{1, j-1} + 0.3L_{1,j-2}+0.1r_{1,j}, & \text{for } j \geq 3
    \end{cases}
\end{align}
where $r_{k,j}\sim(\mathcal{U}[0,\,50]+\sqrt{-1}\cdot\mathcal{U}[-50,\,50])$, with~$\mathcal{U}[a,\,b]$ denoting a~uniform random distribution from $a$ to $b$. For~the second model, we~set
\begin{align}
    L_{2,j} = 0.9L'_{2,j}+0.1r_{2,j},
\end{align}
where $L_{2,j}'$ is a~summation of a~set of sine and cosine terms, each with its frequencies being harmonics of a~set fundamental frequencies. We~set the~cycle corresponding to the~fundamental frequency to be one day. We~then set the~third model to be
\begin{equation}\label{eq:L3model}
    L_{3,j}=\frac{1}{2}(L_{1,j}+L_{2,j})
\end{equation}
as~a~hybrid of both the~auto-regression and the~cyclic~behaviors.

\subsection{Time Series Prediction for Studied Data~Sets} 
\label{subsec:predictDataSet}

In this part, we~develop the~time-series prediction solutions for our studied data sets using the~candidate models described in~Section~\ref{sec:timeSeriesPrediction}. {For the~in-field-data cases, the~SNR data were measured by the~PLMs every 15 min spanning an~overall time period from 17 to 21 months, so that between $N=45,000$ to $N=65,000$ data samples were available to us. For~the synthetic-data case, we~generated $N=63,702$ data samples, which is the~same size as the~two in-field collected data sets with the~documented events. These~values are  sufficiently large so that the~performances of the~trained time-series predictors do not improve with the~further increase of $N$. We~set the~window size to $w=20$, as~we observed from our numerical study that this window size generally results in~a~favorable prediction performance with a~moderate model complexity. For~the subcarrier grouping, we~observed eight notches in~the~frequency band considered for the~PLC data transmission for the~in-field data set. Hence,  setting $n_\textsf{SB}=9$ follows quite naturally as discussed in~Section~\ref{subsec:groupSNR}. This~number also provides for sufficient averaging of per-subcarrier SNRs to obtain~a~smooth time series for the~average SNR in~each subcarrier group.}

\subsubsection{ARIMA}
We consider ARIMA models for all combinations of $p, d, q$, where $0 \leq p, d, q \leq 2$, which is known to be sufficient for most practical time-series prediction tasks~{(}\cite{box2011time}{, Ch.~6)}. Discarding the~case of $p=q=0$, we~investigate a~total of 24 candidate ARIMA~models.

\subsubsection{L2Boost}
\textls[-35]{We choose the~hyper-parameter, $k_\text{total}$, which represents the~total number of iterations as  $k_\text{total}=50,\,100,\,200$. We~make the~choice considering that for smaller values of  $k_\text{total}$, the~resultant trained model has~a~lesser representation power but also a~lower risk of~over-fitting.}

\subsubsection{ANN}

Given the~small input size to the~NN, i.e.,~the window size $w$, we~consider a~simple architecture with one hidden layer with eight neurons for the~FFNN and the~LSTM models. For~the FFNN and LSTM, we~use the~sigmoid function and hyperbolic tangent as the~activation functions for the~hidden layer, respectively. The~purpose of the~activation function for the~hidden layer is to implement a~non-linear transform so that non-linear relationship between the~output and the~input can be learned by the~ANN. 

{\subsubsection{Results}}
Our aim is to develop a~time-series predictor that can predict future values as accurately as possible when the~system is operated under normal conditions, i.e.,~without anomalies. Thereby, an~anomaly produces a~pronounced deviation between the~actual value and the~predicted one. Therefore, in~this part of the~study, the~training and testing data for the~synthetic data sets only contain~the~SNR values when the~cable is under normal operating conditions. For~the data from  field measurements, we~stipulate that most of the~data were collected when the~cables were operated under the~normal condition with only occasional values corresponding to anomalous~conditions. 

{We use $n_{\text{tr}}=0.8\,N$} and the~remaining samples for testing the~performance of the~time series predictor. The~performance of our chosen set of time-series predictors are shown in~Table~\ref{table:mse}, where the~results are presented for the~SNR of the~first stabilizer batch. Exp$_\textsf{MV}$ and Exp$_\textsf{LV}$ are the~two  in-field collected (or experimental) data sets for the~MV network and the~LV network for which instances of network anomalies have been recorded.  {The MV link consists of a~$376$~m three-core cable, and~the LV link comprises a four-core cable over~$69.4$~m long} 
Syn$_{1}$, Syn$_{2}$, Syn$_{3}$ are the~synthetic data sets generated using the~three LI models $L_{k,j}$ for $k=1,\,2,\,3$. For~brevity, we~only present selective results for ARIMA models. From~Table~\ref{table:mse}, we~can observe that FFNN, LSTM, L2Boost and some ARIMA models match or improve the~performance over the~baseline setting. Moreover, the~LSTM model shows the~best performance across the~data sets that we~have investigated, supporting its suitability to time-series prediction tasks. Similar results were obtained for other subcarrier~groups.

\begin{table}[H]
	\centering
	\caption{Normalized RMSE performance of time-series prediction for the~average SNR of the~first stabilizer~batch.} \label{table:mse}
	\newcolumntype{C}{>{\centering\arraybackslash}X}
\begin{tabularx}{\textwidth}{CCCCCC}
\toprule
		\textbf{Data Set} & \textbf{Exp}\boldmath{$_\textsf{MV}$} & \textbf{Exp}\boldmath{$_\textsf{LV}$} & \textbf{Syn}\boldmath{$_{1}$} & \textbf{Syn}\boldmath{$_{2}$} & \textbf{Syn}\boldmath{$_{3}$}\\
		\midrule
		ARIMA$(2,0,1)$ & $43.7\%$ & $38.5\%$ & $52.2\%$ & $61.4\%$ & $20.1\%$ \\
		ARIMA$(2,0,2)$ & $42.9\%$ & $38.5\%$ & $52.2\%$ & $89.9\%$ & $25.2\%$ \\
		\midrule
		ARIMA$(0,1,1)$ & $42.9\%$ & $38.9\%$ & $53.1\%$ & $63.1\%$ & $20.4\%$ \\
		ARIMA$(0,1,2)$ & $42.9\%$ & $38.9\%$ & $53.1\%$ & $62.6\%$ & $20.1\%$ \\
		\midrule				
		L2Boost$(100)$ & $40.2\%$ & $40.5\%$ & $52.2\%$ & $54.8\%$ & $19.2\%$ \\
		L2Boost$(50)$   & $42.0\%$ & $41.7\%$ & $53.1\%$ & $55.6\%$ & $20.5\%$ \\
		\midrule
		FFNN                 & $39.4\%$ & $39.3\%$ & $52.5\%$ & $53.5\%$ & $18.0\%$ \\
		LSTM                 & $39.4\%$ & $38.0\%$ & $52.5\%$ & $51.8\%$ & $17.7\%$ \\
		\midrule
		Baseline             & $43.7\%$ & $40.5\%$ & $56.7\%$ & $70.5\%$ & $20.2\%$ \\
		\bottomrule		
	\end{tabularx} 
\end{table}

We also note from Table~\ref{table:mse} that the~performance of the~baseline model is often fairly close to those from other time-series prediction models. {This is highlighted in~Table~\ref{table:mse2}, which shows the~results for the~ARIMA(2, 0, 1) and the~baseline models for all subcarrier groups. }Therefore, since the~baseline predictor does not require any training and presents no additional computational complexity (see~\eqref{eq:baseline}), the~anomaly detector can begin prediction with this technique until sufficient samples are collected over the~operation to use other predictors that require a~meaningful set of training data. {From the~results in~Table~\ref{table:mse2} we~also observe a~larger variation in~the~prediction accuracy across subcarrier groups for the~in-field compared to the~synthetic data. We~attribute this to the~same LI models and noise PSD used across the~subcarrier groups, which is likely not the~case regarding~the~in-field data.}

\begin{table}[H]
	{
\centering
	\caption{\textls[-35]{{Normalized RMSE} performance of time-series prediction for the~average SNR of all stabilizer batches using the~ARIMA(2, 0, 1) and the~baseline predictors. Results are shown as ARIMA(2, 0, 1)/Baseline.}} \label{table:mse2}

	\newcolumntype{C}{>{\centering\arraybackslash}X}
\begin{tabularx}{\textwidth}{CCCCCC}
\toprule
		\textbf{Data Set} & \textbf{Exp}\boldmath{$_\textsf{MV}$} & \textbf{Exp}\boldmath{$_\textsf{LV}$} & \textbf{Syn}\boldmath{$_{1}$} & \textbf{Syn}\boldmath{$_{2}$} & \textbf{Syn}\boldmath{$_{3}$}\\
		\midrule
		Group 1 & $43.7\%$/$43.7\%$ & $38.5\%$/$40.5\%$ & $52.3\%$/$56.7\%$ & $61.4\%$/$70.5\%$  & $20.1\%$/$20.2\%$ \\
		Group 2 & $39.7\%$/$41.6\%$ & $11.6\%$/$12.4\%$ & $52.3\%$/$57.1\%$ & $60.3\%$/$69.1\%$  & $19.9\%$/$20.9\%$ \\
		Group 3 & $38.4\%$/$39.3\%$ & $30.7\%$/$34.0\%$  & $52.1\%$/$56.9\%$ & $58.5\%$/$67.1\%$ & $19.5\%$/$20.5\%$  \\
		Group 4 & $38.5\%$/$40.3\%$ & $57.4\%$/$65.9\%$ & $52.1\%$/$56.9\%$  & $56.5\%$ /$64.6\%$ & $19.5\%$/$20.4\%$ \\
		Group 5 & $10.4\%$/$10.5\%$ & $38.5\%$/$40.9\%$ & $52.0\%$/$56.8\%$ & $53.9\%$/$61.4\%$ & $19.5\%$/$20.5\%$  \\
		Group 6 & $38.2\%$/$38.9\%$ & $42.6\%$/$48.9\%$ & $52.0\%$/$56.8\%$ & $52.1\%$ /$59.3\%$ & $19.5\%$/$20.5\%$\\
		Group 7   & $17.2\%$/$17.3\%$ & $48.5\%$/$55.7\%$ & $52.0\%$/ $56.8\%$& $51.5\%$/$58.6\%$ & $19.6\%/20.5\%$
       \\
		Group 8   & $17.4\%$/$17.5\%$ & $26.9\%$/$30.2\%$ & $52.0\%$/$56.8\%$ & $51.4\%$/$58.6\%$ & $19.6\%$/$20.5\%$ \\
		Group 9  & $11.4\%$/$10.8\%$  & $41.8\%$/$46.8\%$ & $52.0\%$/$56.8\%$ & $51.4\%$/$58.5\%$ & $19.6\%$/$20.6\%$ \\
		\bottomrule		
	\end{tabularx} }
\end{table}

\subsection{Anomaly Detection for Studied Data~Sets} 
\label{subsec:detectDataSet}

In this section, we~develop and test our anomaly detector for the~studied data sets. According to the~discussion in~Section~\ref{subsec:synthesizeSNR}, we~approximate the~prediction errors for the~average SNR values as~a~multivariate Gaussian distribution, with~a dimension of nine since we~have nine stabilizer batches in~total. We~then calculate $D_\text{MA}^2$ using~\eqref{def_dma} and use~\eqref{eq_threshold} for the~anomaly detection with varying $p_\text{FA}$ and $\kappa=n_\text{SB}=9$.

\begin{figure}[H]
    	\subfloat[][]{\includegraphics[width=7cm]{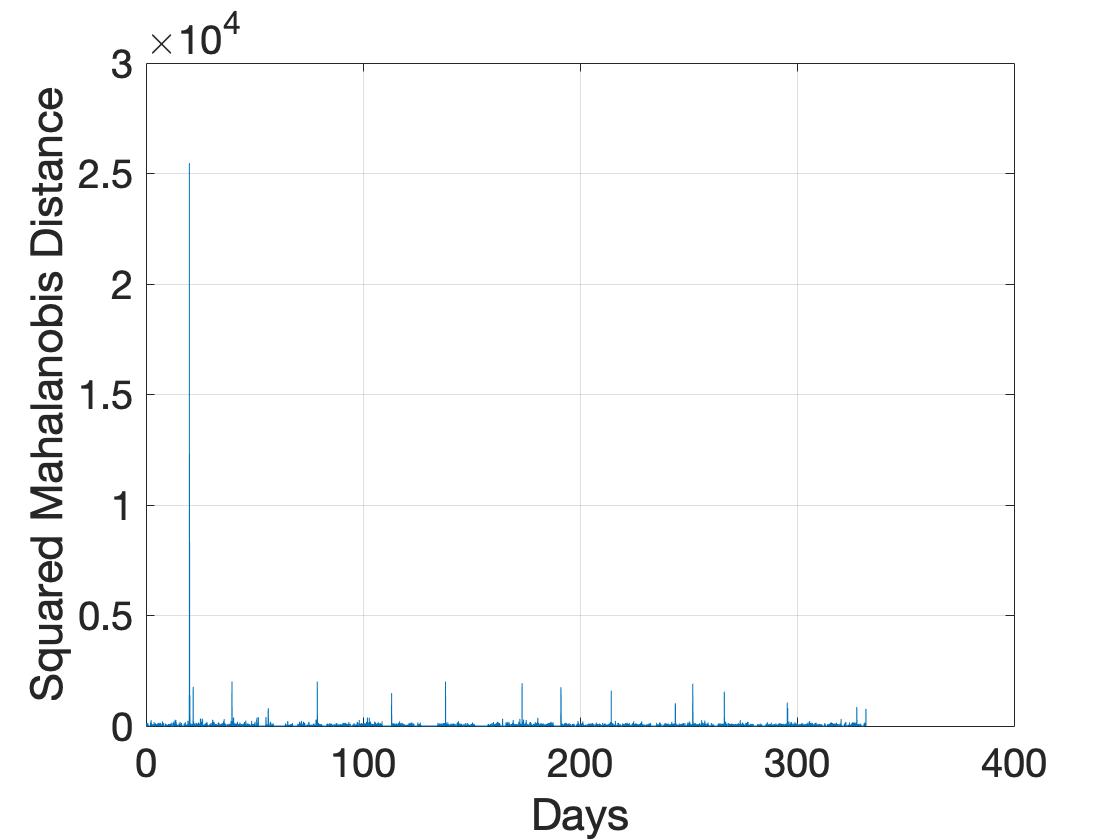} \label{fig:Future_to_Present_MV2}}
    	\subfloat[][]{\includegraphics[width=7cm]{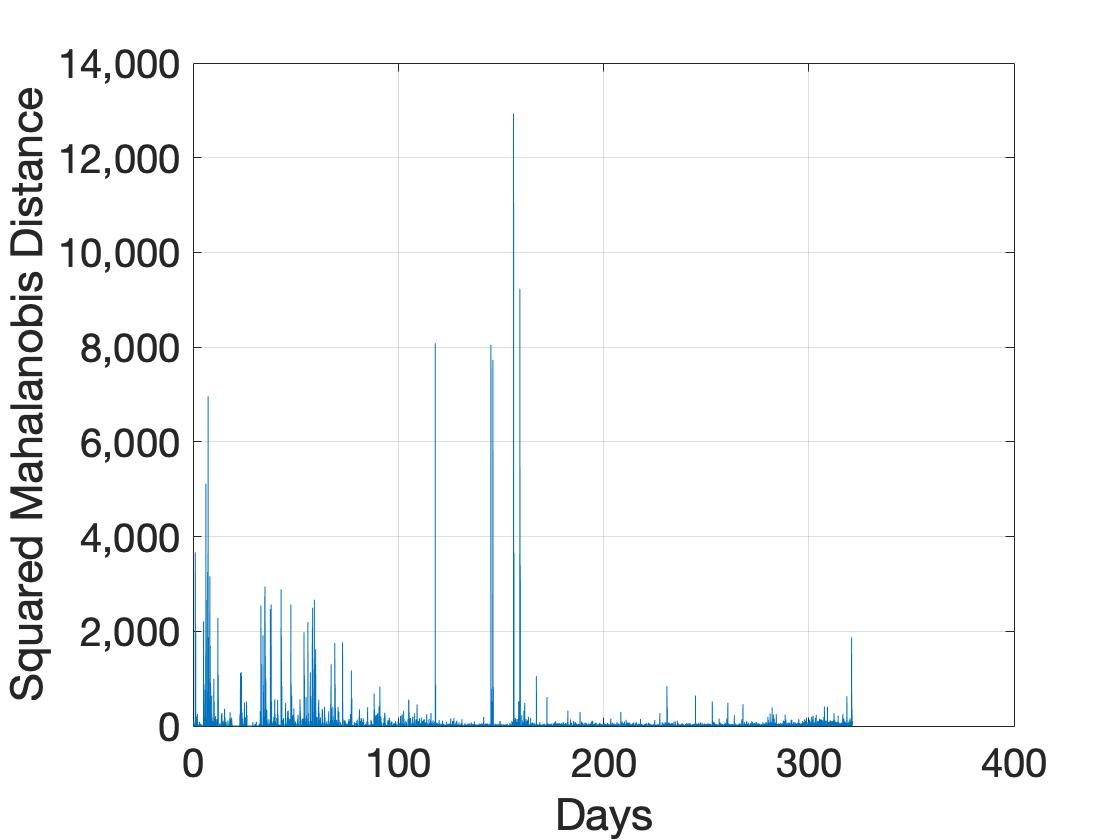} \label{fig:Future_to_Present_LV32}}
    	\caption{\textls[-35]{SMD for the~in-field collected data using ARIMA$(2,1,1)$. {(\textbf{a}) Data Set Exp}$_\textsf{MV}$; (\textbf{b}) Data Set Exp$_\textsf{LV}$.}}
	 \label{fig:Future_to_Present}
      \end{figure}

\begin{figure}[H]
    	\subfloat[][]{\includegraphics[width=7cm]{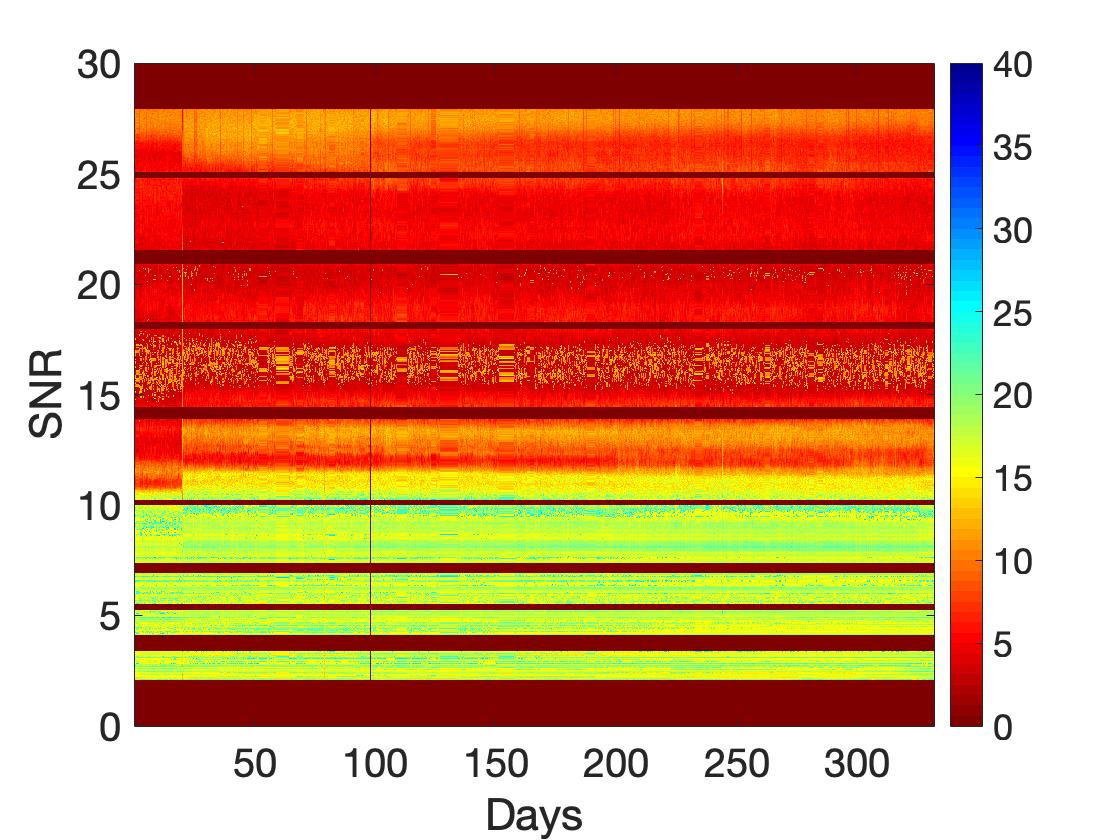} \label{fig:SNR_MV2}}
    	\subfloat[][]{\includegraphics[width=7cm]{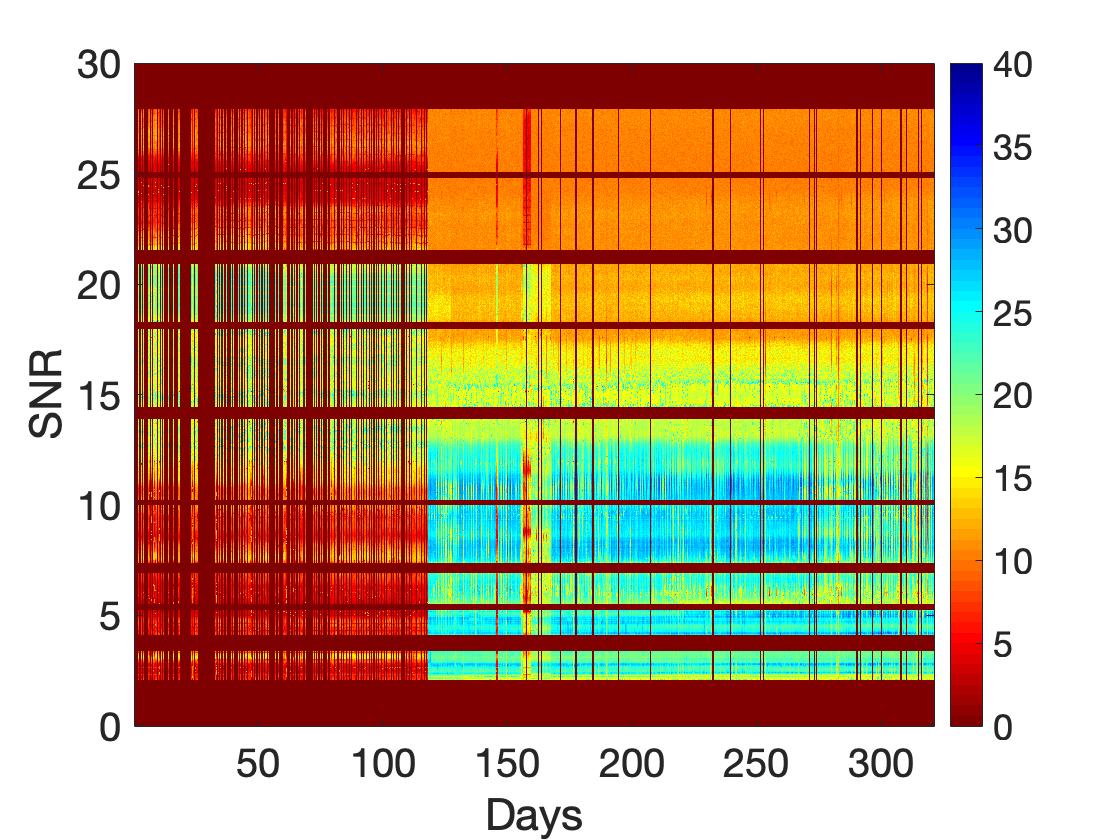} \label{fig:SNR_LV32}}
    	\caption{SNR color map for the~in-field collected~data.  {(\textbf{a}) Data Set Exp}$_\textsf{MV}$; (\textbf{b}) Data Set Exp$_\textsf{LV}$.}
	 \label{fig:SNR}
\end{figure}

The only available recorded anomaly events for the~in-field data in~\cite{hopfer2020nutzen}, are the~switching operations at the~$20$th day in~the~data set Exp$_\textsf{MV}$ and the~fuse failure at the~$156$th day in~the~data set Exp$_\textsf{LV}$. For~each stabilizer batch, we~compute the~average SNR data and calculate the~SMDs based on~the~prediction errors. The~results for data set Exp$_\textsf{MV}$ and data set Exp$_\textsf{LV}$ are shown in~Figure~\ref{fig:Future_to_Present}a and Figure~\ref{fig:Future_to_Present}b, respectively. The~two documented events are clearly seen in~these two figures as notable spikes. To~relate this result with the~observed raw data, we~present the~SNR color maps for the~two data sets in~Figure~\ref{fig:SNR}a,b. It~is also clearly noticeable from the~figures that there are multiple (undocumented) anomalies in~the~LV data in~Figure~\ref{fig:SNR}b, which are rightly represented as notable spikes in~the~SMD plot of Figure~\ref{fig:Future_to_Present}b. The~higher rate of the~indicated abnormal events in~LV networks in~comparison with MV networks can be attributed to the~increased presence of interference and higher disturbance levels in~an LV~network.

\unskip

While the~two documented events in~the~in-field data provided us the~opportunity to test the~performance of our solution using real-world data, the~exercise does not provide a~comprehensive evaluation of our method, especially for different types of cable anomalies and operation under various load types and load changes. To~this end, we~use synthetic training and testing data sets obtained from the~network and LI models constructed as explained in~Section~\ref{subsec:dataSet}. This~provides us the~flexibility to choose a~variety of load and anomaly types to investigate the~robustness of our~method.

We identify three main categories of anomalies, similar to those in~\cite{passerini2019smart}, which are, concentrated faults, distributed faults (DFs), and~abnormal termination impedance changes. We~emulate a~concentrated fault by inserting a~fault resistance $r_\text{f}$ between a~pair of conductors at the~fault point. Such a~line--line fault is the~most common among all types of hard faults~\cite{saha2013unsymmetrical}. This~process can also be extended by placing a~fault impedance $r_\text{f}$ between each pair of conductors to emulate a~symmetrical fault. To~emulate a~DF, we~increase the~per-unit-length (PUL) resistance of the~conductor and the~PUL conductance of the~insulation materials over a~section of the~cable that is affected by this degradation. For~many types of DF, the~conductors have a~deteriorated conductance and the~insulation material has degraded insulation property~\cite{densley2001ageing}, which we~emulate by this process. Finally, to~emulate the~abnormal termination impedance changes, for~our synthetic generators, we~change from one LI model to another, among~the three that we~use, over~a period of time, e.g.,~one hour for four~samples.

\begin{figure}[H]
    	\subfloat[][]{\includegraphics[width=7cm]{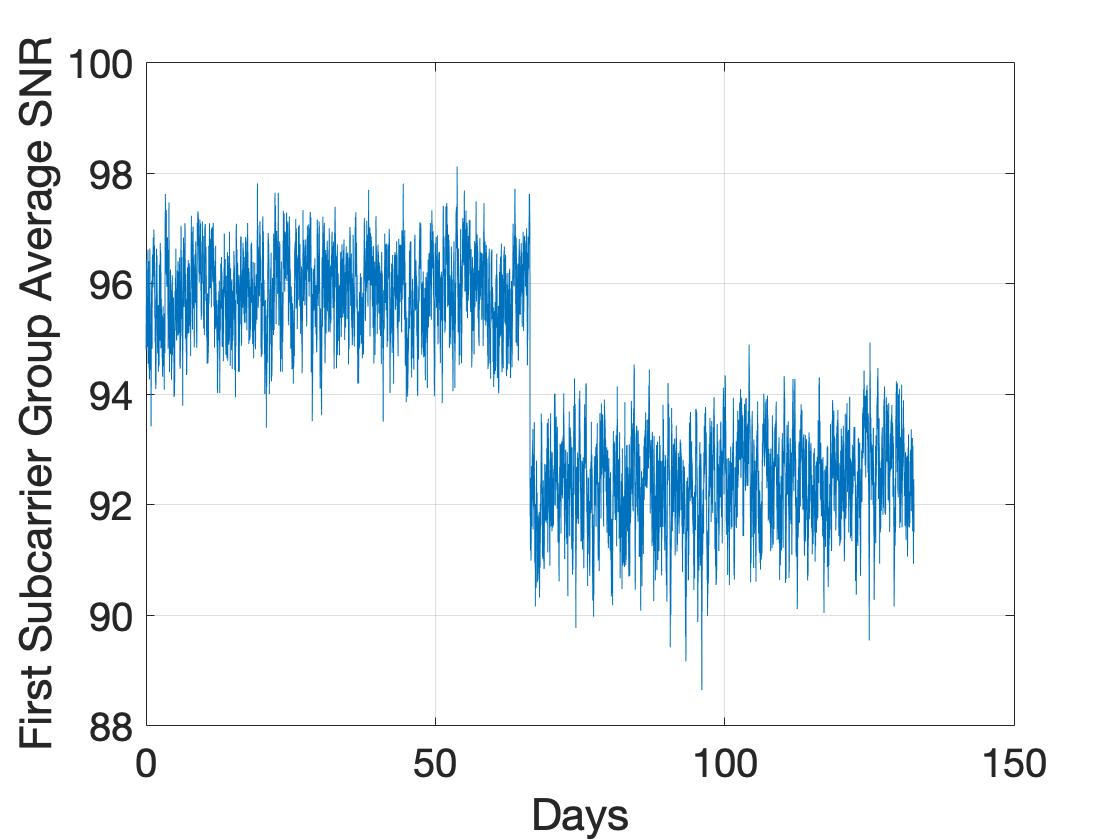} \label{fig:SNR_HF}}
    	\subfloat[][]{\includegraphics[width=7cm]{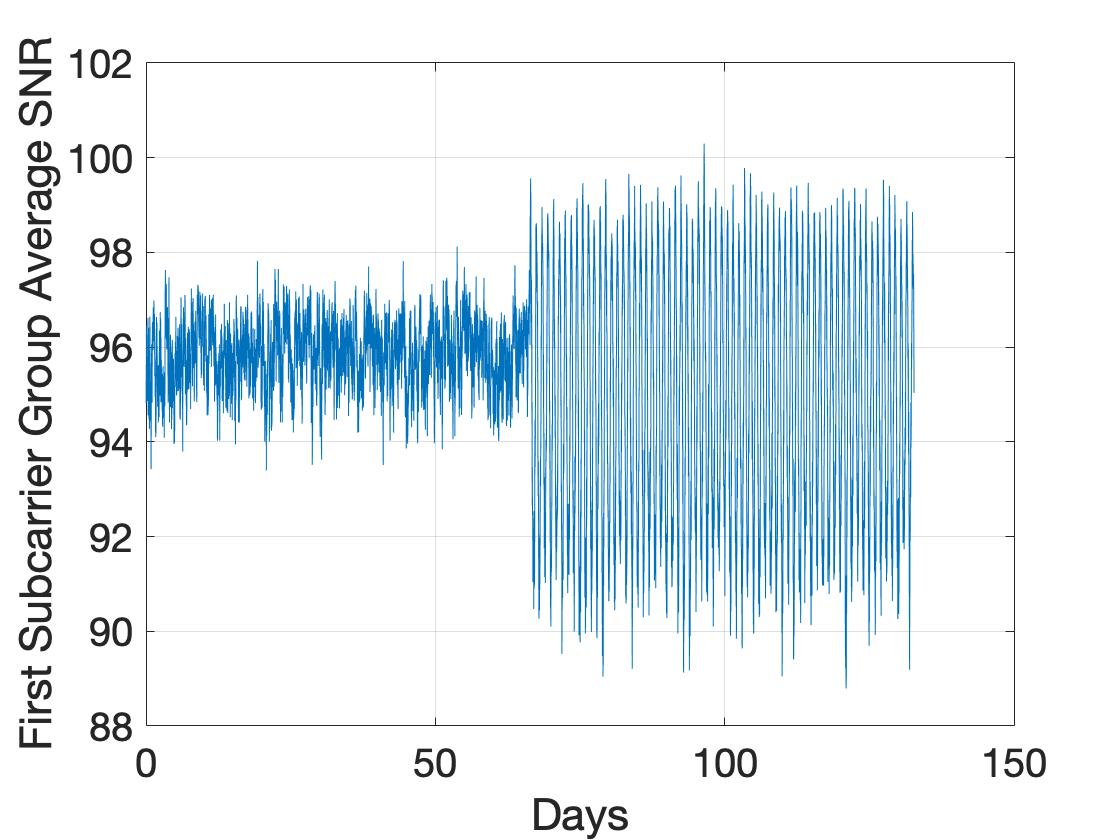} \label{fig:SNR_CL}}
    	\caption{\textls[-25]{Average SNR for the~batch $i=1$. {(\textbf{a}) Concentrated Fault; (\textbf{b}) Termination Impedance Change}.}}
	 \label{fig:SNR_HFCL}
\end{figure}

We first present the~results of the~change in~SNR values with the~introduction of a~concentrated fault. We~introduce a~fault impedance $r_\text{f}=100~\Omega$ between a~pair of conductors at a~location that is $100~\text{m}$ from the~{PLM transmitter, i.e.,~PLM-1 in~Figure~\ref{fig:PLCTnetwork}}. We~show the~impact of this in~Figure~\ref{fig:SNR_HFCL} by contrasting the~average SNR change of one stabilizer batch for the~condition of concentrated fault in~Figure~\ref{fig:SNR_HFCL}a with LI model 1 and a~termination impedance change from LI model~1 to LI model~3 in~Figure~\ref{fig:SNR_HFCL}b. It~is clearly visible that these variations cause a~significant, noticeable, and~distinctive change in~the~measured SNR values. As~a result, we~focus on~the~more challenging case of DF in~the~following.

We introduce three different types of DFs, a~slight DF, a~mild DF, and~a medium DF. We~emulate each of these three conditions by increasing the~PUL serial resistance and shunt conductance of the~cable by $10\%$, $20\%$, and~$60\%$, respectively, to~emulate different extents of cable degradation~\cite{furse2003frequency}. We~introduce the~DF over a~$300$~m section of the~cable with the~starting point of the~faulty section being at a~distance of $100~\text{m}$ away from { PLM-1}.

\begin{figure}[H]
    	\subfloat[][]{\includegraphics[width=7cm]{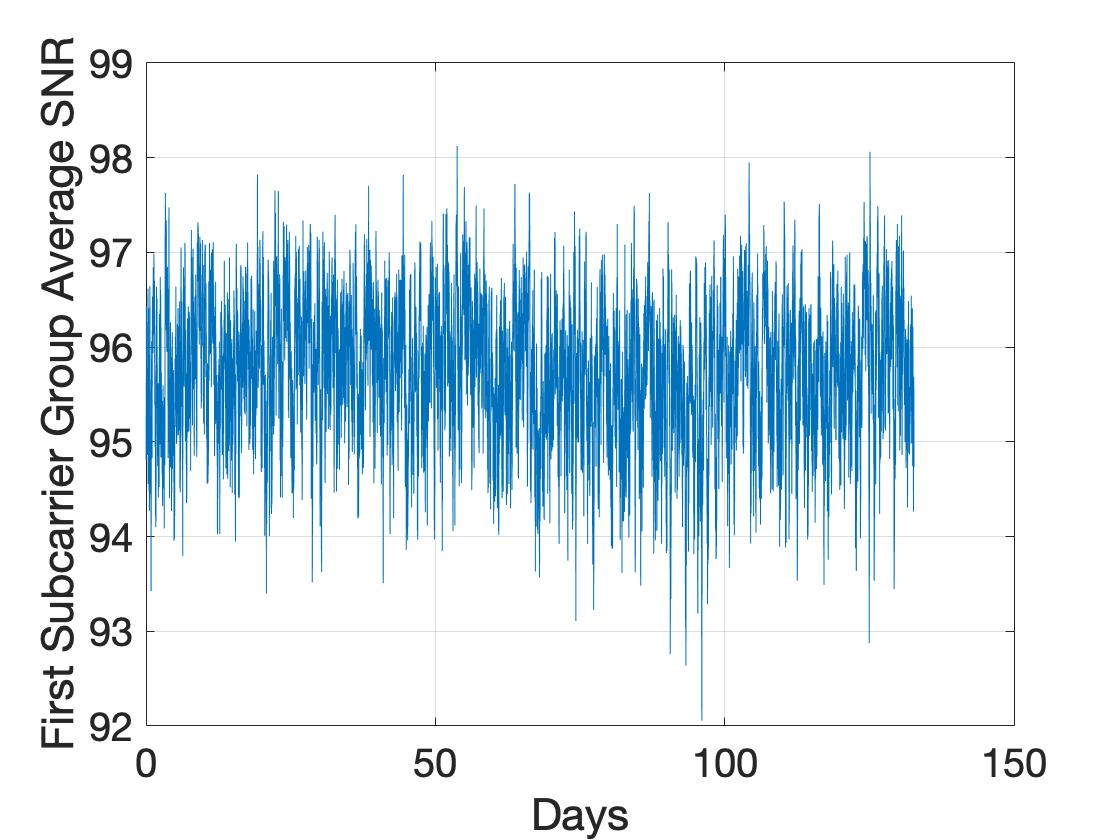} \label{fig:SNR_DF_mild}}
    	\subfloat[][]{\includegraphics[width=7cm]{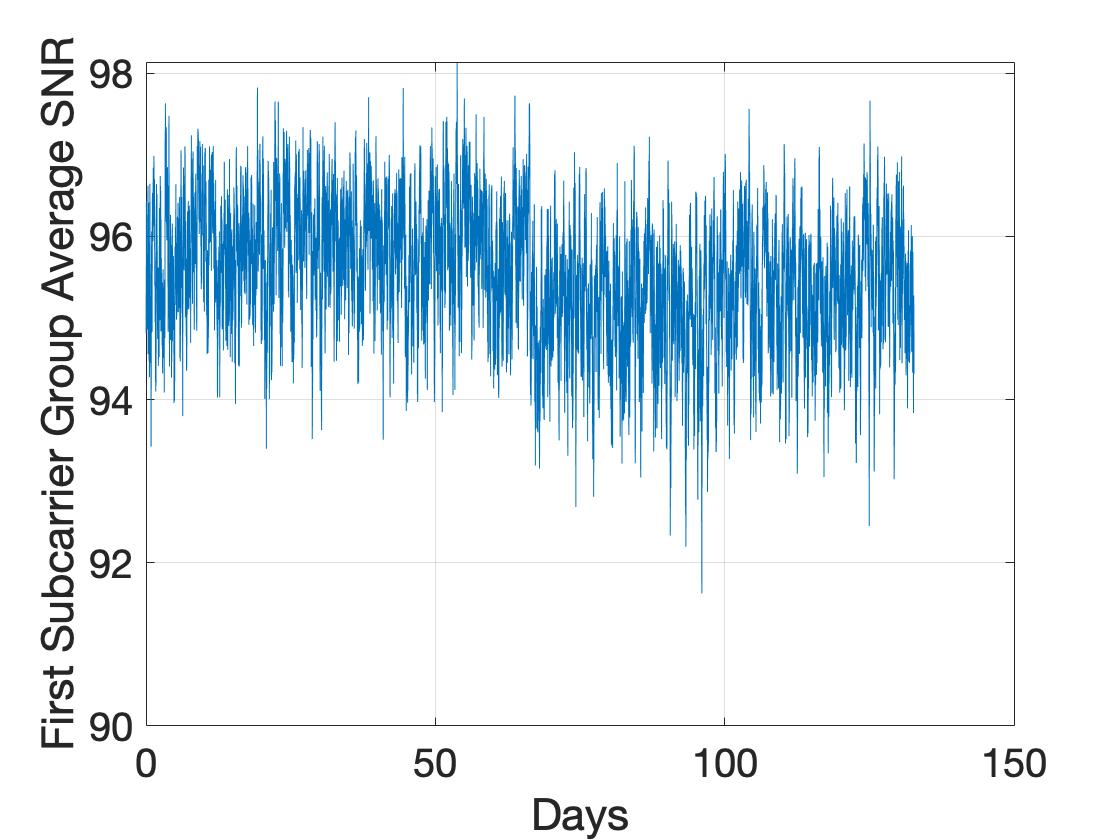} \label{fig:SNR_DF_medium}}
    	\caption{Average SNR with mild and medium DFs for the~stabilizer batch $i=1$ with LI model~$1$. {(\textbf{a})~Mild DF; (\textbf{b}) Medium DF}.}
	 \label{fig:SNR_DF}
\end{figure}

\begin{figure}[H]
    	\subfloat[][]{\includegraphics[width=7cm]{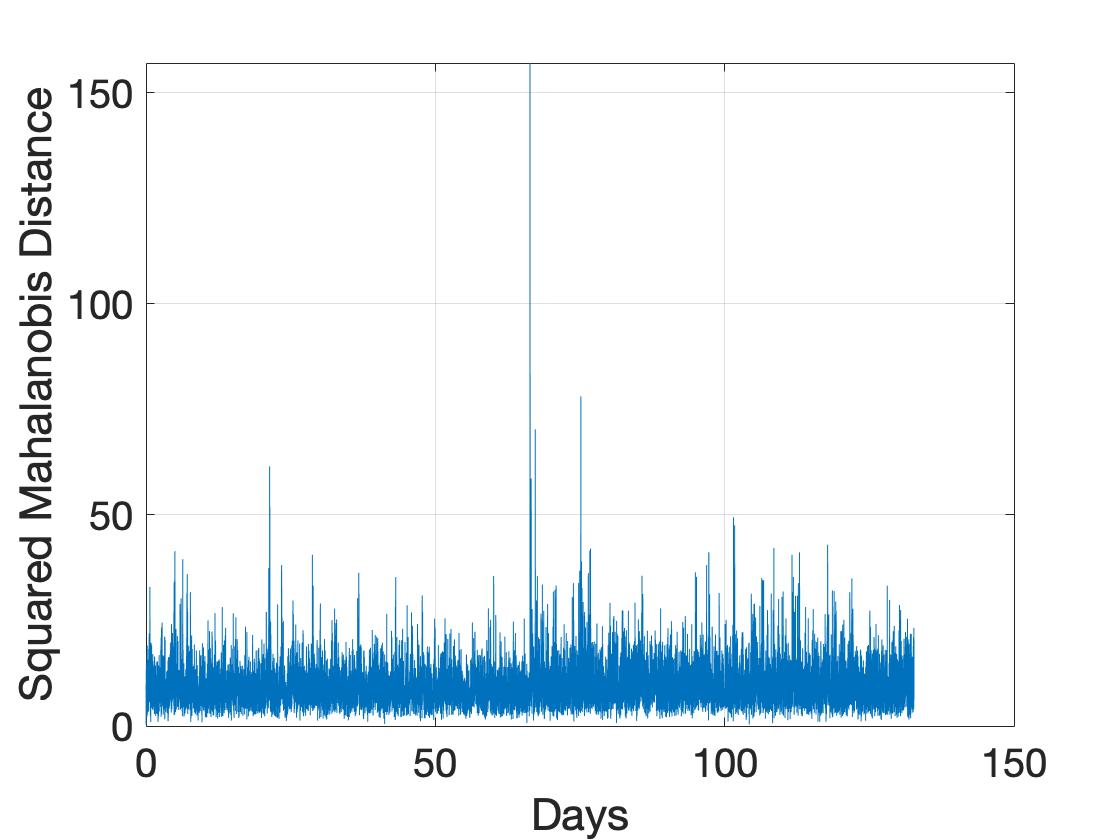} \label{fig:SMD_DF_mild}}
    	\subfloat[][]{\includegraphics[width=7cm]{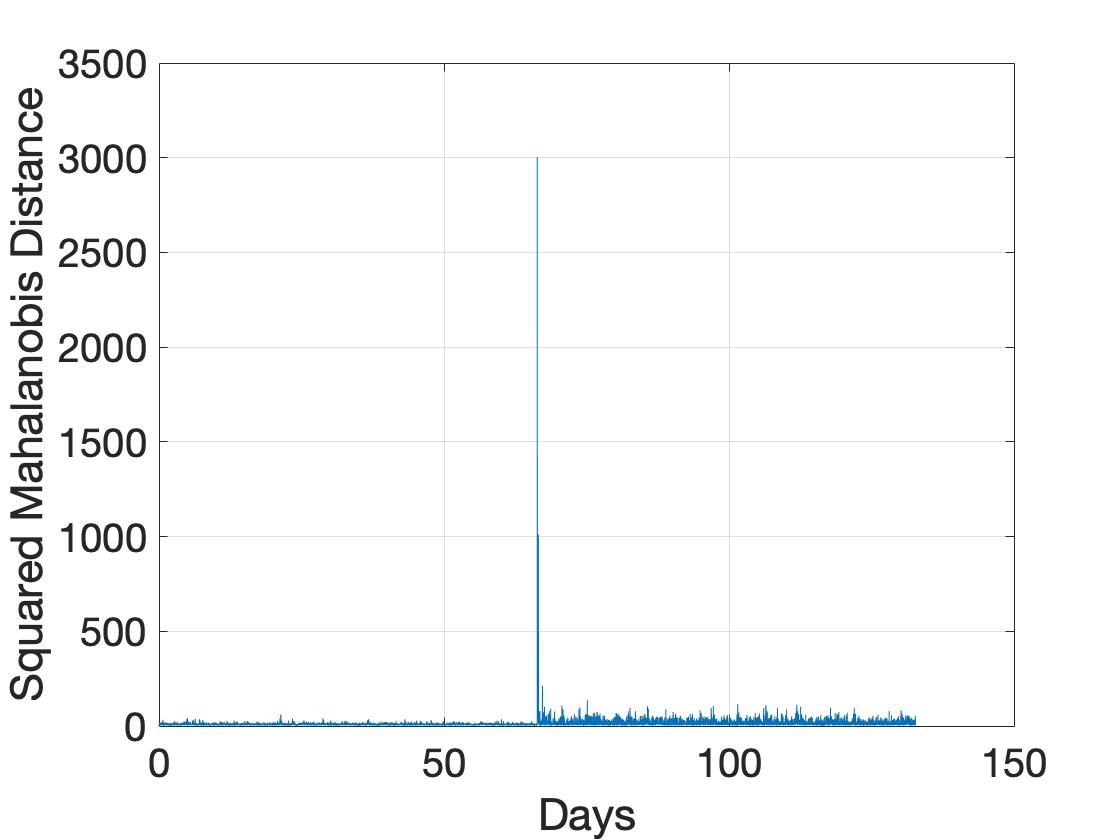} \label{fig:SMD_DF_medium}}
    	\caption{SMD for mild and medium DFs whose SNRs are presented in~Figure~\ref{fig:SNR_DF}. {(\textbf{a})~Mild DF; (\textbf{b})~Medium DF}.}
	 \label{fig:SMD_DF}
\end{figure}

The average SNR values of the~first stabilizer batch over time, as~shown in~Figure~\ref{fig:SNR_DF}a,b, signify that detecting a~DF is more challenging than a~hard fault. 
We employ our anomaly detection procedure, and~accordingly compute the~SMDs, as~illustrated in~Figure~\ref{fig:SMD_DF}, where the~faulty events are indicated as distinctive spikes in~the~middle. We~then determine the~anomaly detection thresholds with an~FA rate $p_\text{FA}$  either theoretically using~\eqref{eq_threshold} or empirically through the~training data. For~the empirical determination, we~sort $|D_\text{MA}^2|$ for the~training data prediction difference in~the~descending order as $d_i$ from $d_1$ to $d_{(n_\text{tr}-w)}$. We~then compute the~threshold as
\begin{equation}\label{eq:threshold_empirical}
	T_{\text{r}}(p_\text{FA}) = d_{\left\lfloor {p_\text{FA}\cdot(n_\text{tr}-w)} \right\rfloor},
\end{equation}
where $\left\lfloor{\cdot}\right\rfloor$ is the~floor~function.

\begin{figure}[H]
	\includegraphics[width=12cm]{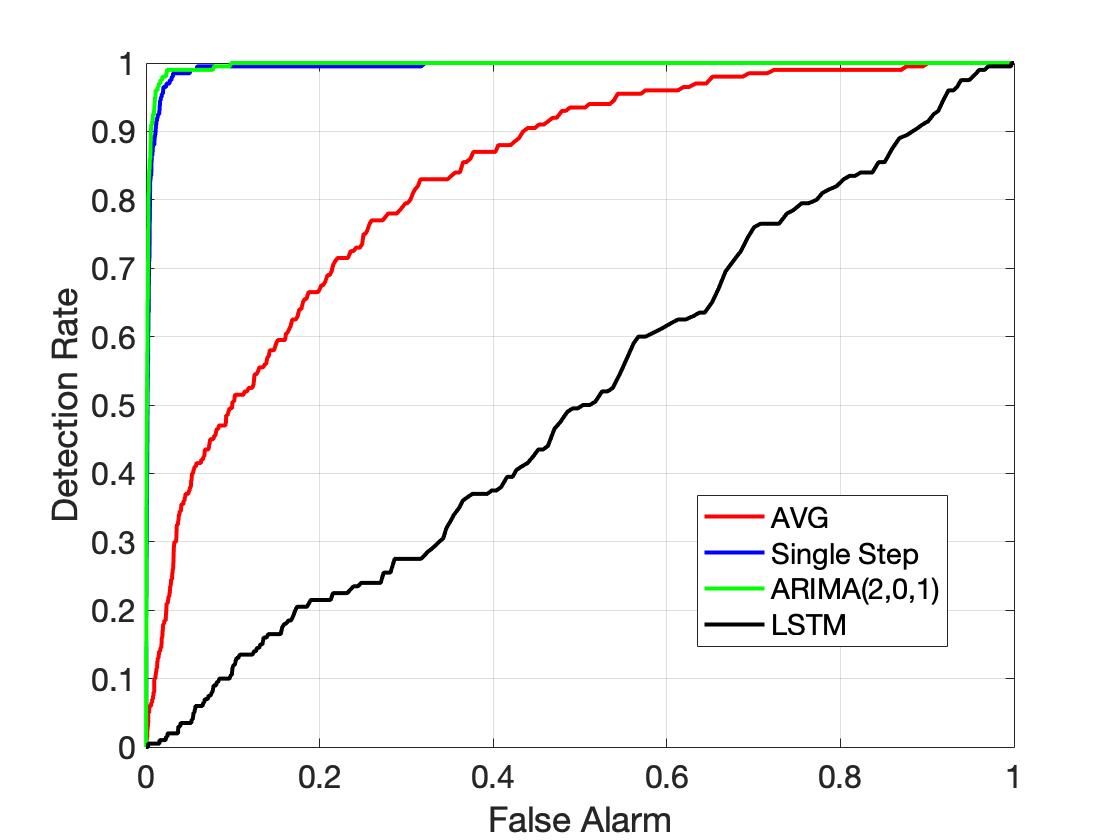}
	\caption{ROC for the~generated test cases with slight DF.}
	\label{fig:ROC_curve}
\end{figure}

Choosing the~threshold involves a~trade-off between the~detection rate $p_\text{DT}$, i.e.,~the probability that an~anomaly can be successfully detected, and~$p_\text{FA}$, i.e.,~the probability that a~normal condition is identified as an~anomaly. An~increase in~detection rate is typically accompanied by higher FA rates. We~show this behavior in~the~receiver operating characteristic (ROC) curve for our anomaly detection solution in~Figure~\ref{fig:ROC_curve}. Since the~performance of our method for the~cases of mild and medium DFs are nearly ideal for all candidate forecasting choices, we~only present ROC behaviors for the~more challenging case of slight DF. We~generate $100$ different test cases, where in~each case, we~introduce a~slight DF in~the~middle of the~time series. The~blue single-step (SS) curve in~Figure~\ref{fig:ROC_curve} is the~baseline prediction method in~\eqref{eq:baseline}, and~AVG is an~alternative trivial prediction scheme that uses the~average of the~training data as the~predicted value at all times. We~observe from \linebreak Figure~\ref{fig:ROC_curve} that ARIMA and baseline predictors provide the~best detection performance, as~also evidenced in~Table~\ref{table:mse} for prediction performance. However, anomaly detectors using LSTM or other data driven time-series predictors demonstrate worse performance than even the~AVG predictor for the~case of slight DF. We~observe that data driven time-series predictors, including LSTM, FFNN and L2Boost, have good prediction performance both before and after the~slight DF is introduced. This~shows that they adapt better to the~case of faulty cable condition. Such generalization ability to unobserved data with a~slight difference from the~training data is a~detriment to anomaly detection as it~does not produce a~distinct change of the~prediction error after the~slight DF is introduced. For~more distinct DFs however, e.g.,~mild and medium DFs, anomaly predictors using data driven time-series predictors and those using the~classical ARIMA models have matched~performance.

For an~FA rate of $p_\text{FA}=1\%$, we~obtain~the~threshold as $T_{\text{r}}(p_\text{FA})=21.67$ theoretically using~\eqref{eq_threshold}, or~empirically using the~training data and~\eqref{eq:threshold_empirical} as $T_{\text{r}}(p_\text{FA})=23.91$ as an~alternative. For~the generated test cases, the~threshold to achieve an~FA rate of $p_\text{FA}=1\%$ is $T_{\text{r}}(p_\text{FA})=23.75$, which is very close to the~threshold determined theoretically using~\eqref{eq_threshold} or empirically using the~training data. This~shows that both theoretical and empirical approaches are viable methodologies to determine the~threshold $T_{\text{r}}(p_\text{FA})$.

\section{Supplementary~Evaluation} 
\label{sec:discuss}
In this section, we~further investigate the~suitability of our proposed solution in~practical scenarios. In~particular, we~address two challenges faced in~practice, which are the~lack of available data for training and the~identification of cable anomalies that are gradual in~nature, such as an~incipient~fault.

\subsection{Robustness~Test}
\label{subsec:robustness}

\begin{figure}[H]
    	\subfloat[][]{\includegraphics[width=7cm]{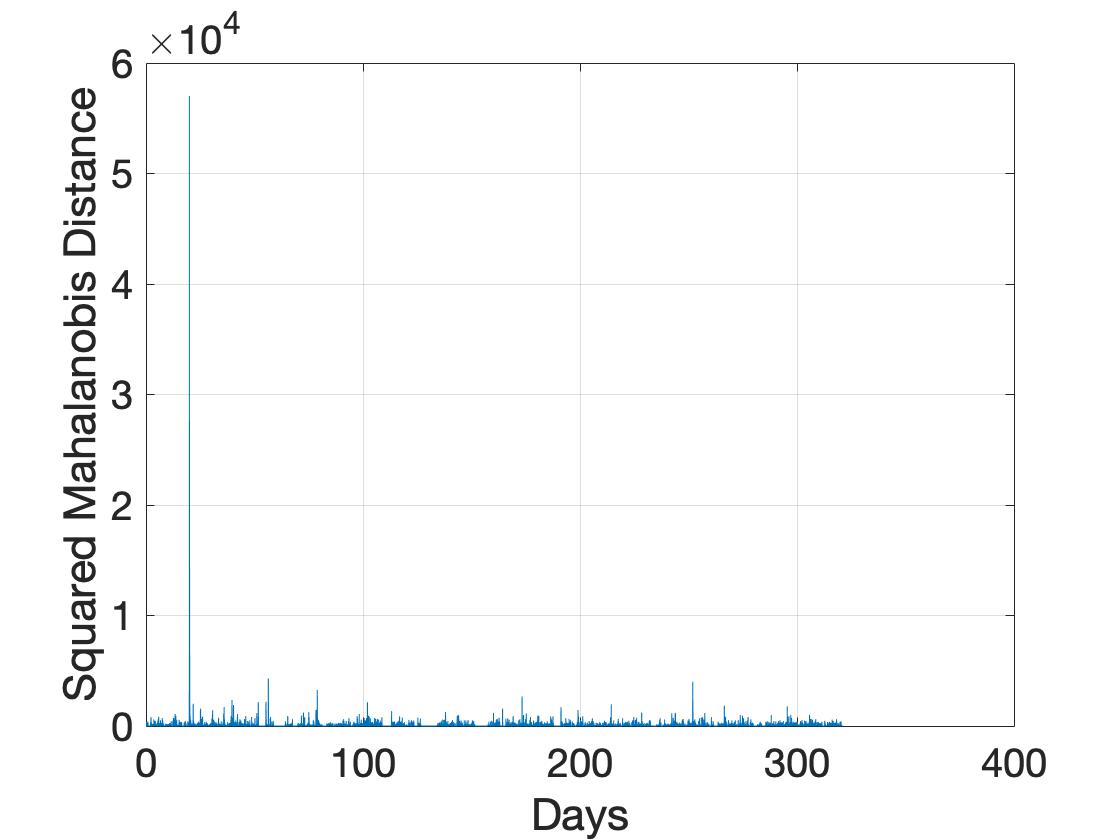} \label{fig:Utility_to_Utility}}
    	\subfloat[][]{\includegraphics[width=7cm]{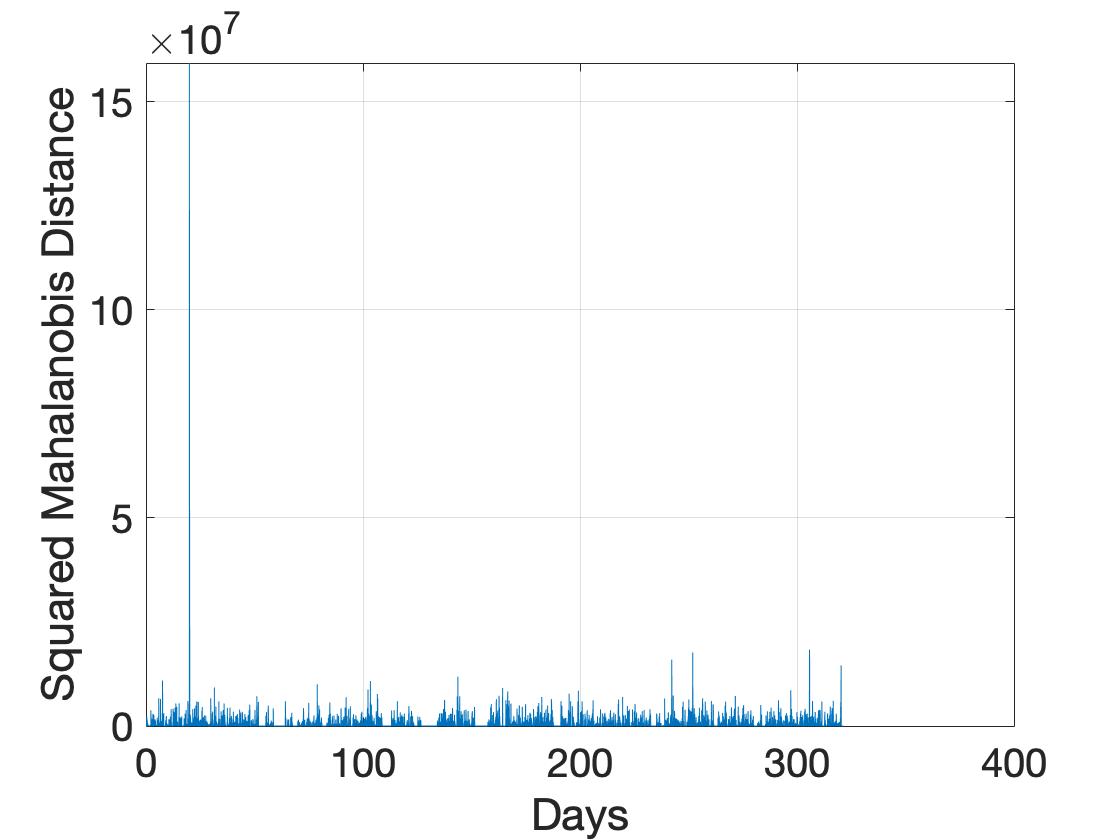} \label{fig:Synthetic_to_Utility}}
    	\caption{SMD for the~in-field collected SNRs from the~data set Exp$_\textsf{MV}$. {(\textbf{a})~Trained Using Exp}$_\textsf{MV2}$; (\textbf{b})~Trained Using Synthetic Data.}
	 \label{fig:SMD_DF_Robust}
\end{figure}

Our evaluation campaign in~Section~\ref{sec:caseStudy} involved using historical SNR time-series data for training and prediction. This~type of data collection is suitable in~fixed asset monitoring. However, we~investigate the~suitability of using our solution as~a~dynamic diagnostics technique, where a~machine is trained to detect anomalies on one type of a~network and required to function on another type. This~expands the~scope of our proposed solution to make it~more universally applicable, where, e.g.,~the SNR data from one pair of transmitter and receiver can be used to detect anomalies in~networks operating in~a~different portion of the~grid. A~likely more beneficial use-case is to train~the~machine using synthetic data extracted from a~\textit{best-guess} estimate of the~network-under-test and to use it~in~a~real-world network to detect cable anomalies. We~conduct both these investigations and present the~performance results in~Figure~\ref{fig:SMD_DF_Robust}a,b. 

In the~first evaluation, we~train~the~machine using SNR data extracted from the~dataset Exp$_\textsf{MV2}$, another MV experimental dataset from~\cite{hopfer2020nutzen}, and~test it~over the~data set collected in~a~different portion of the~MV network, i.e.,~dataset Exp$_\textsf{MV}$. The~result in~Figure~\ref{fig:SMD_DF_Robust}a shows a~clearly discernible spike in~the~SMD plot, which is easily detectable by our anomaly detector with little/no FA. The~adjacent Figure~\ref{fig:SMD_DF_Robust}b demonstrates that training the~network with synthetic data, which were generated using $L_3$ according to the~procedure explained in~Section~\ref{subsec:dataSet}, and~testing it~with the~in-field collected data set Exp$_\textsf{MV}$ is also able to detect network anomalies. The~results from Figure~\ref{fig:SMD_DF_Robust}a,b indicate the~robustness of our solution to variations between training and application data.
\vspace{-.1in}

\subsection{Incipient~Fault}
\label{subsec:incipient}

\begin{figure}[H]
	\includegraphics[width=9cm]{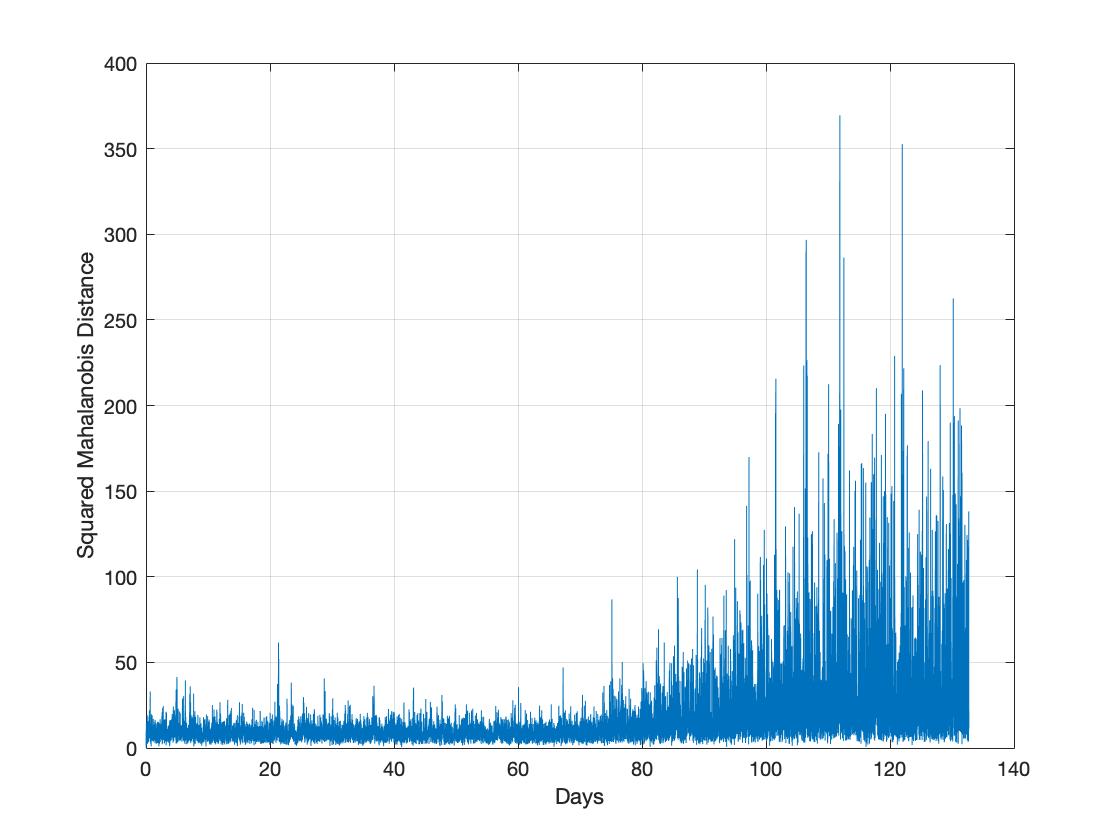}
	\caption{{SMD for} detecting an~incipient fault.}
	\label{fig:SMD_IF}
\end{figure}

Our investigations in~Section~\ref{sec:caseStudy} considered faults that are abrupt, i.e.,~occurring to their full extent at one instant of time. However, the~cable may also be susceptible to an~incipient fault, which is introduced gradually over time. We~emulate such a~condition by generating a~$132$-day time sequence, where the~incipient fault begins to develop on~the~$66$th day. We~quantify the~severity of the~fault after the~$66$th day by $\gamma(t) \propto t$, where $t$ is time in~seconds. We~{increase the~PUL serial resistance and PUL shunt conductance by a~factor of $\gamma(t)$ between $\gamma(t)=0$ on~the~$66$th day to $\gamma(t)=2$ on~the~$132$nd day.} We place the~incipient fault on a~cable section of $300$~m whose starting point is $100$~m from the~transmitter PLC modem, {i.e., PLM-1,} and use $L_1$ to generate our synthetic SNR data. We~train~the~predictor using normal operating conditions, i.e.,~without the~incipient fault, and~then use ARIMA$(2,1,1)$ for time-series forecasting. The~resultant SMD for the~generated incipient fault case is shown in~Figure~\ref{fig:SMD_IF}. The~SMD plot shows spikes indicating a~fault from the~$66$th day onward and whose magnitude increases as time progresses. Naturally, the~choice of the~threshold determines how quickly an~incipient fault can be detected and what the~FA rate is that is sacrificed in~the~process. This~decision would be made based on~the~operating~scenario.

\section{Conclusions}
\label{sec:conclusion}
We have designed a~\textit{first-of-its-kind} PLC-based universal cable anomaly detector as~a~smart grid sensing solution using time-series forecasting and statistical test of prediction errors. Our low-cost solution repurposed PLC modems as sensors to also enable monitoring of the~grid system to ensure its smooth operation and improve its resilience by reusing the~channel state information inherently estimated by the~modems. Our method, which combines forecasting with the~post-processing of prediction errors based on Mahalanobis distance, produces a~robust cable anomaly detection performance. Further, our solution is applicable across various network conditions and can operate without prior domain knowledge of the~anomaly, network topology, type of~cable, or~load~conditions.

\section*{Acknowledgment}
\label{sec:ack}

This work was supported by funding from the Natural Sciences and Engineering Research Council of Canada (NSERC). The authors would also like to thank Dr. Nikolai Hopfer from the University of Wuppertal, Germany, and Power Plus Communications AG (PPC), Germany, for making the experimental data available and assisting with the data analysis. The experimental data was collected in a research project supported by the German Federal Ministry of Education and Research [grant number 03EK3540B].

\bibliographystyle{ieeetr}
\bibliography{TestBibTex}{}

\end{document}